\documentclass[preprint]{aastex}

\usepackage{amsmath,amsfonts}

\usepackage[utf8]{inputenc}
\usepackage{natbib}
\usepackage{threeparttable}
\usepackage{url}
\usepackage{lscape}
\usepackage{soul}
\title{On the interpretation of the Fermi GBM transient observed
in coincidence with LIGO Gravitational Wave Event GW150914}
\shortauthors{Connaughton et al.}

\author{V.~Connaughton\altaffilmark{*,1}, 
E.~Burns\altaffilmark{2,+}, 
A.~Goldstein\altaffilmark{1},
L.~Blackburn\altaffilmark{3,4}, 
M.~S.~Briggs\altaffilmark{5,6}, 
N.~Christensen\altaffilmark{7,8},
C.~M.~Hui\altaffilmark{9},
D.~Kocevski\altaffilmark{9},
T.~Littenberg\altaffilmark{9},
J.~E.~McEnery\altaffilmark{2},
J.~Racusin\altaffilmark{2},
P.~Shawhan\altaffilmark{10},
J.~Veitch\altaffilmark{11},
C.~A.~Wilson-Hodge\altaffilmark{9},
P.~N.~Bhat\altaffilmark{6},
E. Bissaldi\altaffilmark{12,13},
W.~Cleveland\altaffilmark{1},
M.~M.~Giles\altaffilmark{14},
M.~H.~Gibby\altaffilmark{14},
A.~von~Kienlin\altaffilmark{15},
R.~M.~Kippen\altaffilmark{16},
S.~McBreen\altaffilmark{17},
C.~A.~Meegan\altaffilmark{6},
W.~S.~Paciesas\altaffilmark{1},
R.~D.~Preece\altaffilmark{5},
O.~J.~Roberts\altaffilmark{1},
M.~Stanbro\altaffilmark{5}, 
P.~Veres\altaffilmark{6}
}

\altaffiltext{*}{Email: valerie@nasa.gov}
\altaffiltext{1}{Universities Space Research Association, 320 Sparkman Dr., Huntsville, AL 35805, USA}
\altaffiltext{2}{NASA Goddard Space Flight Center, Greenbelt, MD 20771, USA}
\altaffiltext{+}{NASA Postdoctoral Fellow USRA}
\altaffiltext{3}{Harvard-Smithsonian Center for Astrophysics, 60 Garden St., Cambridge, MA 02138, USA}
\altaffiltext{4}{LIGO, Massachusetts Institute of Technology, Cambridge, MA 02139, USA}
\altaffiltext{5}{Dept. of Space Science, University of Alabama in Huntsville, 320 Sparkman Dr., Huntsville, AL 35805, USA}
\altaffiltext{6}{CSPAR, University of Alabama in Huntsville, 320 Sparkman Dr., Huntsville, AL 35805, USA}
\altaffiltext{7}{Physics and Astronomy, Carleton College, MN, USA 55057}
\altaffiltext{8}{Artemis, Universit\'e C\^ote d’Azur, Observatoire C\^ote d’Azur, CNRS, CS 34229, F-06304 Nice Cedex 4, France}
\altaffiltext{9}{Astrophysics Office, ST12, NASA/Marshall Space Flight Center, Huntsville, AL 35812, USA}
\altaffiltext{10}{Department of Physics, University of Maryland, College Park, MD, USA 20742}
\altaffiltext{11}{University of Glasgow, Glasgow G12 8QQ, United Kingdom}
\altaffiltext{12}{Istituto Nazionale di Fisica Nucleare, Sezione di Bari, 70126 Bari, Italy}
\altaffiltext{13}{Dipartimento di Fisica, Politecnico di Bari, 70125, Bari, Italy}
\altaffiltext{14}{Jacobs Technology, Inc., Huntsville, AL, USA}
\altaffiltext{15}{Max-Planck-Institut f\"ur extraterrestrische Physik, Giessenbachstrasse 1, 85748 Garching, Germany}
\altaffiltext{16}{Los Alamos National Laboratory, NM 87545, USA}
\altaffiltext{17}{School of Physics, University College Dublin, Belfield, Stillorgan Road, Dublin 4, Ireland}

\date{}

\newcommand{\fermi}{{\it Fermi }}
\newcommand{\Fermi}{{\it Fermi}}
\newcommand{\rmfit}{{\it rmfit }}
\newcommand{\Rmfit}{{\it rmfit}}

\begin{abstract}
The weak transient detected by the \fermi Gamma-ray Burst Monitor (GBM) 
0.4~s after GW150914 has generated much speculation regarding 
its possible association with the black-hole binary merger.
Investigation of the GBM data by Connaughton et al. (2016)
revealed a source location consistent with GW150914 and a
spectrum consistent with a weak, short Gamma-Ray Burst. 

Greiner et al. (2016) present an alternative technique
for fitting background-limited data in the low-count regime, and call into question the spectral analysis and the significance of
 the detection of GW150914-GBM presented in Connaughton et al. (2016).

The spectral analysis of Connaughton et al. (2016) 
is not subject to the limitations 
of the low-count regime noted by Greiner et al. (2016). 
We find Greiner et al. (2016)
used an inconsistent source position and did not follow the steps taken
in Connaughton et al. (2016) to mitigate the statistical shortcomings
of their software when analyzing this weak event. 
We use the approach of
Greiner et al. (2016) to verify that our original spectral analysis is not biased.

The detection significance of GW150914-GBM is established empirically, 
with a False Alarm Rate (FAR) of $\sim 10^{-4}$~Hz. 
A post-trials False Alarm Probability (FAP)
of $2.2 \times 10^{-3}$ ($2.9 \sigma$) of this transient being associated 
with GW150914 is based on the proximity in time to the GW event of a transient with
 that FAR. 
The FAR and the FAP are unaffected by the spectral analysis that is the focus of
 Greiner et al. (2016).  
\end{abstract}

\begin{document}
\maketitle

\section{Introduction\label{sec:intro}}

With the detection by  
the Laser Interferometer Gravitational-wave Observatory (LIGO; \cite{aligo}) \citep{abbott2016,abbott2016b,abbott2017}
of two highly significant gravitational wave (GW) events and one additional probable GW event during their O1 science run came 
the search for possible electromagnetic counterparts.
%all 
%from merging stellar-mass black holes in binary systems 
A concerted observational follow-up campaign, organized in advance of the first science operation 
period of LIGO, O1, involved dozens of
 ground- and space-based telescopes, including  
the  Gamma-ray Burst Monitor (GBM) on the \fermi Gamma-ray Space Telescope.  
GBM is an all-sky monitor of the transient sky between 8~keV and 40~MeV, consisting of 12 Sodium Iodide (NaI) 
scintillators sensitive below 1~MeV and 2 Bismuth Germanate (BGO) scintillators sensitive above 200~keV. 
\cite{meegan2009} provide a comprehensive description of GBM and we list only the GBM capabilities that are salient
to the work presented here.
The NaI detectors have different orientations that together cover the whole sky with highest sensitivity along the observatory
pointing axis, and falling off at very large angles
to the boresight of the observatory. 
By examining the relative count rates measured in the 12 detectors the arrival direction of any detected signal
can be reconstructed with an accuracy ranging from 
tens to hundreds of square degrees on the sky, depending on the intensity of the event.  
The BGO detectors have a more
omnidirectional response and contribute mostly to the spectral analysis of transients above 200~keV.
GBM's participation in the follow-up campaigns to GW events during O1 contributes broad sky coverage, coarse but 
useful source localization capability, high duty cycle, and energy
coverage that provides good sensitivity to short Gamma-Ray Bursts (sGRBs).
A summary of the follow-up observations to the first GW event, GW150914, is presented in \cite{singer2016}, with 
more observational details in \cite{singer2016b}.  

Because the most likely progenitor for a sGRB is the merger of compact objects in a binary system involving at
least one neutron star, and because the number of sGRBs detected in the LIGO horizon is low \citep{siellez2014} 
-- zero so far with measured redshift -- 
the GBM team has deployed offline searches of the GBM data for 
sGRBs too weak to trigger GBM on-board.  These offline searches are described in more detail in 
\cite{vc2016} (VC+16). An untargeted offline search of the GBM data 
yields around 80 new candidates per year (Briggs et al., in preparation), tripling the GBM sGRB detection rate. 
A targeted search of the GBM data designed
specifically to look for counterparts to GW events is even more sensitive.
Details of the targeted offline search deployed during O1 are provided in \cite{blackburn2015} 
and VC+16. It can be summarized as a search over the whole sky, coherently
combining the data from all 14 GBM detectors (NaI and BGO) 
to test the statistical preference for a source above background. 
The search is performed over a user-specified time window, 
 revealing short-duration candidates typically between 0.256~s to 8.192~s in duration,
with candidates ranked by a Bayesian likelihood statistic.
At each tested sky position the full instrument response is convolved
in turn with three template source spectra.
The likelihood of a source
being present is evaluated with each template. 
The preferred spectral template and the most likely arrival direction for any transient are
the template and sky position that maximize the likelihood.
A high likelihood value implies that the relative rates in the 14 detectors are consistent with a
source coming from that direction.
Conversion of the likelihood into a FAR came originally
from running the search on 
2 months of GBM data \citep{blackburn2013} 
and depends on the preferred template spectrum for each candidate. 
The FAR does not assign meaning to a particular transient --  
it is calculated empirically with respect to the distribution of transients in the GBM data, 
including astrophysical transients and background fluctuations.
There are more soft 
transients in the GBM data than hard, likely associated with galactic sources. 
For a given likelihood value, therefore, a transient that prefers the soft spectrum has a higher FAR than one
preferring the hard spectrum.  
In VC+16 we report the detection with the targeted offline search of the GBM data
of a transient preferring the hard spectral template,
 0.4~s after GW150914, with a FAR of about $10^{-4}$~Hz. 
We verified the conversion of likelihood into FAR established
in \cite{blackburn2013} by running the search on 220~ks of data taken on days around the GW event,
and in \cite{vc2016} we used the very similar FAR from the contemporaneous data in preference to the earlier result. 
%A False Alarm Probability (FAP) can be calculated by considering the FAR and
%the temporal offset of the transient from the GW event in any search window.  
In the absence of theoretical predictions for an electromagnetic signal from a binary black
hole merger, we assume that the closer in time a candidate occurs relative to the GW time, the 
more likely it is that the events are related. 
Assuming the probability of association scales inversely with the relative time to the GW event,
we find a FAP of 0.0022 for a candidate with this FAR. 
%In our earlier work, we 
% assume a linear dependence in time for the probability that the event occurred
%by chance at a given temporal offset from the GW event and calculate a 
Under a more conservative assumption that the relative time probability is uniform over a 60~s search window, the FAP would be 0.028. 
This empirical result emerges from a procedure developed {\it a priori} by running the search
on  months of GBM data and is
independent of and unaffected by subsequent investigation of the data to probe the nature of the
transient.  
 
The targeted search can additionally consider the consistency of possible
GBM sky locations with the source position region derived from the
LIGO data.
This capability was not implemented during O1 because we had not assessed
the effect on the location-dependent likelihood of combining information from
imperfect GBM localizations with LIGO sky regions of a different shape but similar size.
Instead, a uniform sky prior was assumed and the source could come from any direction.
We explored further the localization of the gamma-ray transient using the process employed by the GBM team
in regular operations. This process informed the development of
the targeted search, which retains common features such as its
use of spectral templates and grid searches for likely source arrival directions.
  
\begin{figure}
  \centering
    \includegraphics[width=6in]{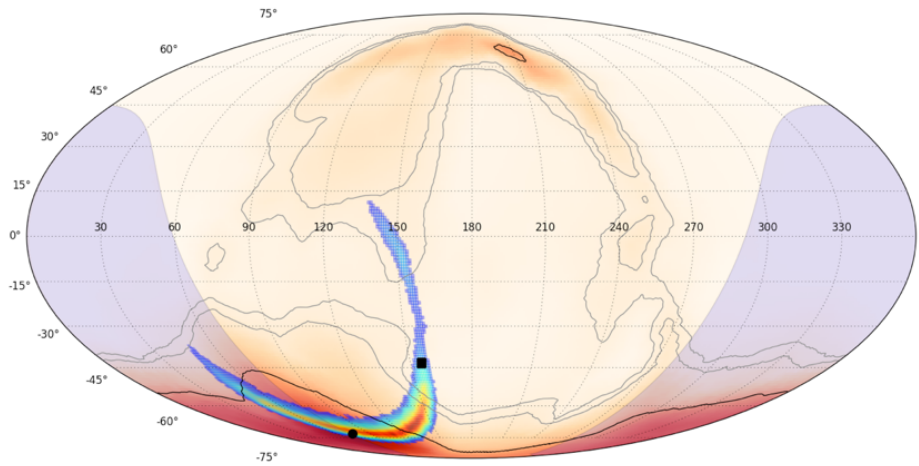}
   \caption{Localization in equatorial coordinates
of GW150914 by LIGO (colored arc) and of
GW150914-GBM by GBM (1, 2, 3 $\sigma$ contours in black, 
reddish shading showing probability gradient).  Most of the probability
for GBM is contained in the southern hemisphere with a slight possibility that
the source came from a mirror point in the north. The most 
likely position from the GBM localization is on the Earth's limb (Earth
shading in purple) and is marked by a black circle.  The position used
in the JG+16 analysis is shown by a black square and lies
outside the $3 \sigma$ GBM localization contour, implying the relative
detector rates are incompatible for a source coming from this position.
\label{fig:gbm_loc}}
\end{figure}

The localization code \citep{connaughton2015} finds the minimum $\chi^2$ on
a $1^\circ$ resolution grid of arrival directions.
The observed background-subtracted rates in the 12 NaI detectors 
are compared with the rates expected from simulations 
of three template spectra using the detector responses
for sources at each point on the grid.  A
statistical uncertainty region is defined by tracing the gradient of $\chi^2$ within the grid around the most 
likely position.  
When applied to GW150914-GBM, this uncertainty region is large, with the 68\% confidence
level region covering 3000 sq degrees, as shown in Figure \ref{fig:gbm_loc}, 
and encompasses large regions of the LIGO localization arc.  
The relative GBM detector rates are incompatible at this confidence level with a source outside the uncertainty region.
The systematic uncertainties for triggered GRBs are
small compared with this large region \citep{connaughton2015} but we have not yet estimated any additional systematic effects
that might apply to the localization of such a weak transient, and we omit systematic uncertainties from these localization
contours.
Consistency is seen between the arrival directions of
the GBM-detected transient and the GW event, albeit with large uncertainties in both instruments.  
Combining the GBM localization
with the LIGO localization region 
and excluding the part of the sky hidden to \fermi by the Earth allows a 2/3 reduction of the LIGO localization region, 
assuming of course the events are related.

In VC+16, we report spectral fits to the GBM data for GW150914-GBM 
performed using the \rmfit spectral analysis 
software package\footnote{\url{http://fermi.gsfc.nasa.gov/ssc/data/analysis/rmfit/}} 
and appropriate detector responses. 
%attempting simple power-law fits in energy as well as power-law fits
%with an exponential cut-off in energy.
We sampled eleven positions along the LIGO arc in Figure \ref{fig:gbm_loc} without
considering whether the positions were consistent with the GBM localization, but excluding source
locations behind the Earth to \Fermi.  
Using data from the NaI and BGO detectors with the smallest angles to the source position, in all cases NaI 5 and BGO 0, 
we were able to obtain a power-law fit to the data for all positions along the LIGO arc,
with an index value of $-1.40^{+0.18}_{-0.24}$ sampled over the arc.  An exponential cut-off fit was possible for 
one source position, in 
approximately 50\% of the fit iterations, 
but the position was excluded as a possible source location
by the GBM localization so that this should be considered an unreliable fit.
The fluence calculated between 10 and 1000~keV, 
obtained by deconvolving the instrument response from positions sampled uniformly over the arc, was  
$2.4^{+1.7}_{-1.0} \times 10^{-7}$~erg~cm$^{-2}$,
among the 40\% weakest short GRBs that trigger GBM. 
The fact that this transient did not trigger GBM is explained by its arrival geometry, at very large angle to the observatory boresight
relative to which the GBM detectors are aligned.  At the most likely source location, the detector with the
smallest angle to the source direction is NaI~5 at $70^\circ$. 
None of the NaI detectors had good, on-axis coverage of the source direction and thus 
many detectors registered the event counts via the back and side. 
None of the detectors individually registered statistically significant increases in the count rate above the background.
Owing to the large angles of the NaI detectors
 to the observatory boresight,
any source signal passes through significant obstacles on the satellite, suffering absorption at the
lower energies and scattering at all energies. 
Because of this viewing
geometry, although the fluence is among the distribution of fluences 
calculated for triggered short GRBs, it is much weaker in count space
than GRBs that trigger GBM. 
%Given that the localization process depends on being able to measure and compare the relative rates in each detector,  
%the low source counts in each detector large localization uncertainties 

\cite{greiner2016} (JG+16) present an alternative spectral fitting technique 
they have developed for fitting 
background-limited data in the low-count regime,  
and call into question the spectral analysis of VC+16 and the significance of the detection of GW150914-GBM in VC+16. 
The spectral analysis of VC+16 is not subject to the limitations of the low-count regime noted by JG+16.
The detection significance of GW150914-GBM is unaffected by the spectral analysis that is the focus of JG+16.
Our findings are summarized below.

\begin{itemize}
\item
JG+16 compare the spectral fit results obtained for simulated source spectra
from the \rmfit software used in VC+16 with those from
their own software package. As the simulated source becomes weaker, they show that
\rmfit overestimates the amplitude of the spectrum  
and their software does not.  
In VC+16 we use 8-channel data to perform the spectral deconvolution and not the finely-binned 128-channel data used in JG+16.
This mitigates the known limitations of \rmfit when calculating fit parameters and their uncertainties in the 
low-count regime.
\item 
As evidence that \rmfit overestimates the amplitude of the spectrum,
JG+16 convolve the fit parameters of VC+16 with a detector response and find the predicted counts
above background exceed the observed counts in that detector.
The fit parameters in VC+16 were obtained by sampling source positions 
over the entire LIGO arc. 
The fit of JG+16 is for a single source direction, where the detector response is nearly
three times greater in effective area than most of the positions on the arc.
When we convolve our fit parameters for this single favorable source position with the appropriate detector
response, we match the observed counts in the detector, implying no \rmfit-induced bias exists in our fit.
\item
We repeat our spectral fits with 8-channel data and
the XSPEC software package using a fitting statistic appropriate for 
the low-count data. We find results very similar to VC+16, showing our
spectral fits using \rmfit are not biased by low-count statistics.  
The amplitude parameters in JG+16 are 
significantly lower and spectral indices softer
(though with very large uncertainties) than those obtained with XSPEC (or \Rmfit). 
\item
Visual inspection of the background levels in JG+16 suggests they are higher than those in VC+16.
These background fits
result in lower background-subtracted
counts in the source interval, hence lower fit amplitudes and fluences in an analysis of the source spectrum.  
We suggest that a different approach to fitting background levels 
is the main difference between the fits in the two papers, 
and not the limitations of the spectral fitting software used in VC+16.
\item
If the background fits of JG+16 are correct,
then the lower fluences they obtain suggest compatibility with non-detection by SPI-ACS from
 all positions on the arc.  JG+16 do not draw this natural conclusion, 
implied by their own spectral analysis, instead concluding that the lower (but non-zero) fluence indicates no source is present. 
\item
JG+16 look for a statistical preference for a source plus background over just background using
data from just two detectors. VC+16 use data from all 14 detectors in the source detection procedure and in
the calculation of its significance and FAR. 
\item
The source position selected by JG+16
for spectral analysis of GW150914-GBM 
is completely excluded by the GBM localization ($> 3\sigma$ level).
Testing for a source at that position will not yield evidence of a source because
the relative rates measured in the GBM detectors are incompatible with a source at that
location. 
\item
The assertion of JG+16 that the spectrum of GW150914-GBM is ``very soft" like a galactic source is not borne out by
their own spectral fits or by cursory visual inspection of the detector count data.  
\item
JG+16 state that the FAR for GW150914-GBM is an optimistic lower limit because the true spectrum of GW150914-GBM is softer than the
template spectrum with which it was discovered.
The FAR is derived empirically and does not require that the spectral templates be good representations of true source spectra.
The search may not be as sensitive to a source if its spectral
templates do not adequately represent the true source spectrum. This will result in a less sensitive search, not an unreliable FAR; 
in fact for an inefficient search we might say the FAR is a pessimistic upper limit.  
\item
The raw count-rate lightcurve included in VC+16 (Figure 7 in appendix C of VC+16)
 shows a signal-to-noise ratio of $6 \sigma$, contradicting a statement in 
 JG+16 that the signal-to-noise ratio of the lightcurve depends on an assumed hard spectral
shape.
\item
The lightcurves in VC+16 are shown only as a visual aid.  
The signal-to-noise ratio of 5.1 in the model-dependent discovery lightcurve does not relate directly
to the post-trials significance of $< 3 \sigma$ calculated in VC+16, 
as suggested by JG+16.  
The post-trials significance of $2.9 \sigma$
reflects an empirical measurement of how likely it is that a transient of the signal size and
 consistency with a point source indicated by the likelihood
 and associated FAR
(whether background or astrophysical) occurs by chance
so close in time to a GW event.  
\end{itemize}

In the following sections we address these issues in more detail.
We do this without challenging the premise of JG+16
that there exist more suitable statistical approaches to the spectral analysis
of weak transients than that used in \Rmfit.

%Although there are no predictions or well-established mechanisms for detectable EM emission from stellar mass binary black hole mergers 
%to guide a search for counterparts in the GBM data, we carried out a methodical search around the time and sky location of the event GW150914,
%which we report in the following section.

\section{Analysis of GW150914-GBM}
\subsection{Approaches to spectral analysis\label{sec:altspec}}

Both the \rmfit spectral fitting package
used in VC+16 and the maximum-likelihood estimation (MLEfit) package in JG+16 
take a forward folding approach to determining the parameters that best fit the data for any model, 
given the instrumental response to a source from a particular direction.
The minimization routine producing the best fit parameters uses a likelihood-based fitting statistic, CSTAT,
but the approaches to modeling the uncertainties in the background differ.  \rmfit uses a $\chi^2$-based
polynomial fit to the background in which the background uncertainties across
the source time interval are obtained from the model.
MLEfit uses the Poisson likelihood-based Cash statistic in the estimation of the background uncertainty in any time bin.
JG+16 show in section 2.3 of their paper, using simulations
of GRBs over a range of source intensities,
that \rmfit and MLEfit yield similar results for bright sources. As the 
simulated source intensity decreases, \rmfit analysis of 128 energy channel GBM data overestimates the fluence of 
the source whereas MLEfit recovers a value closer to the true value. 

%\begin{itemize}
%\item
$\rightarrow$ In VC+16 we use 8-channel data to perform the spectral deconvolution instead of the 128-channel data used in JG+16.
The pitfalls of using \rmfit to analyze low-count data are known -- 
in the routine calculation by the GBM team of the
duration of a GRB, where spectral fits are performed over successive short time intervals, 
 8-channel data are used in preference to 
128-channel data, a practice we follow in VC+16.  
This mitigates the known effect of low-count statistics in calculating fit parameters and their uncertainties. The
energy channel with the fewest source counts -- 13 counts in channel 6 of BGO~0 -- is above
the limit of 10 counts suggested by JG+16 as a lower limit for \rmfit and CSTAT.
%\end{itemize}

JG+16 acknowledge that 8-channel data are used in VC+16, 
but their exploration of statistical techniques is done with 128-channel data, which makes
 the comparison with the analysis in VC+16 difficult.  
%even though it illustrates well the limitations of \rmfit in the low-count regime.  
This exploration continues in section 3.1 of JG+16 with  
spectral fits to the 128-channel data from GW150914-GBM, 
using \rmfit and MLEfit, 
assuming in turn each of the ten source positions on the southern
part of the LIGO arc analyzed in VC+16, 
and recovering both a power-law index and an amplitude as parameters
of the fit. They find that \rmfit returns higher amplitudes, harder power-law indices, and smaller 
parameter uncertainties for all positions on the arc, although the \rmfit values are contained within the
68\% uncertainty regions of the MLEfit parameters.

\subsection{Use of incorrect detector response}

\begin{figure}
   \centering{
    \includegraphics[width=3in]{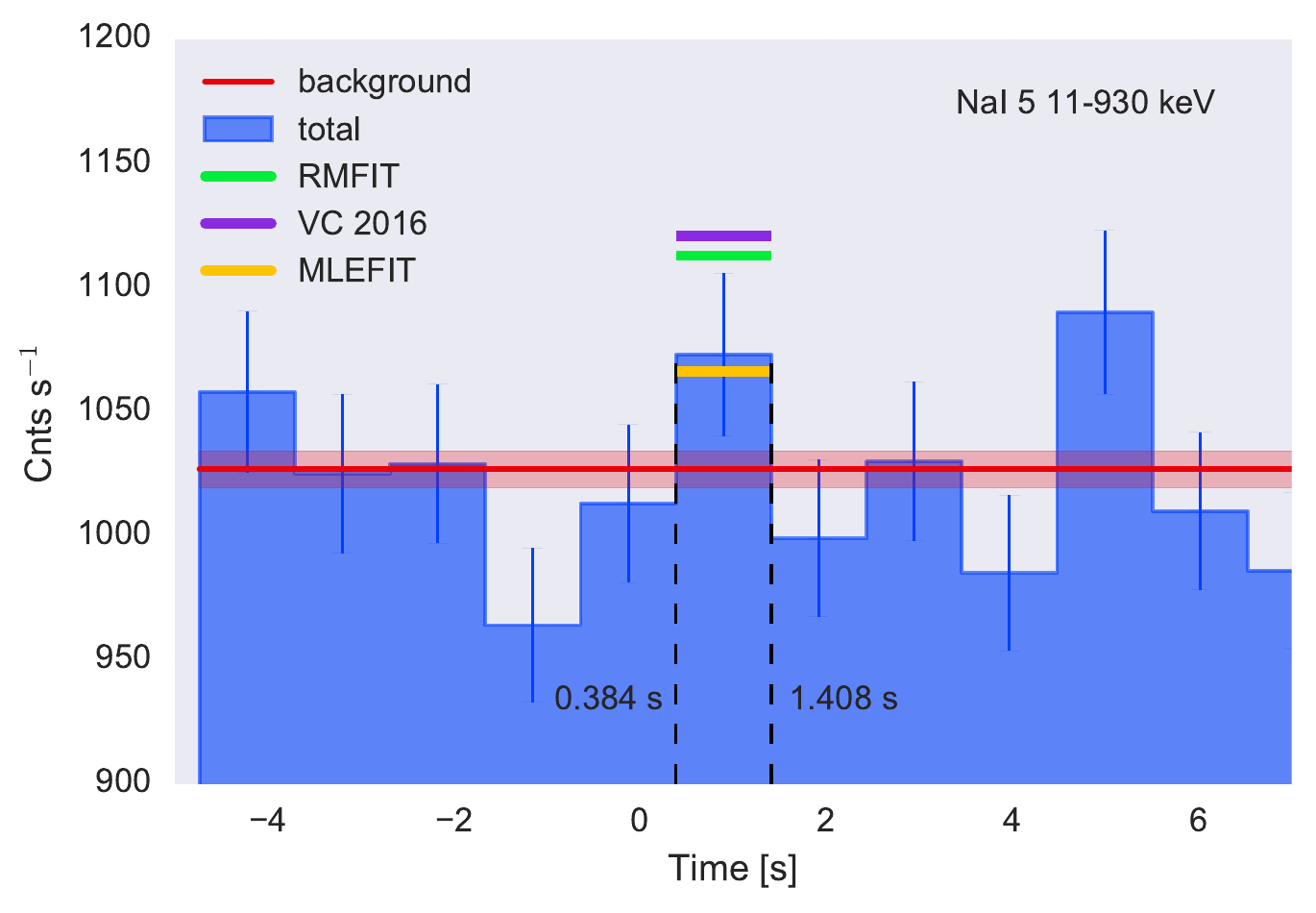}
    \includegraphics[width=3in]{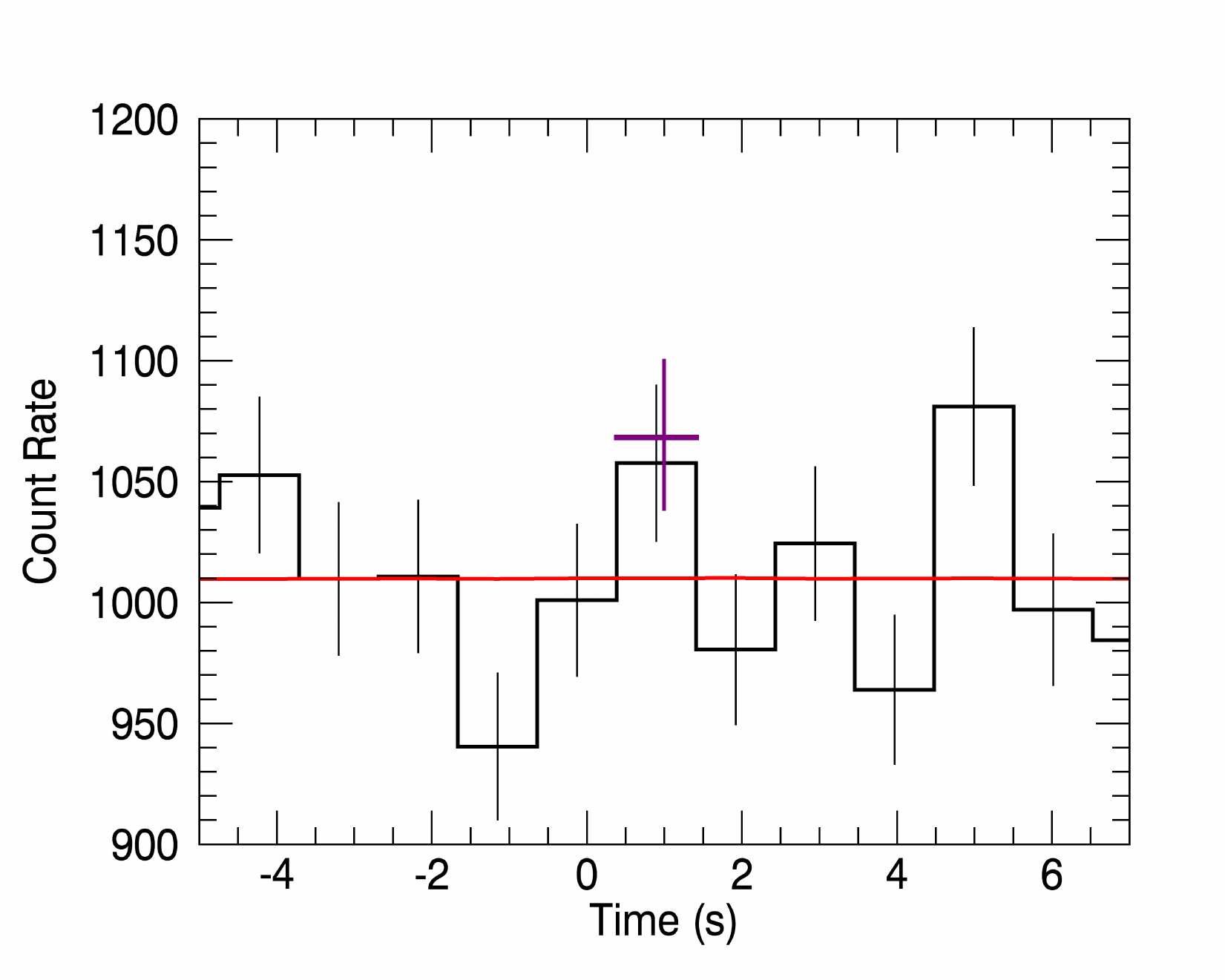}
    \caption{The left panel reproduces Figure 5 of JG+16 which shows the data for NaI~5 near the time of the
GW event (blue histogram). The red band shows the background level from their fit and the colored data points show the
predicted count rates obtained by convolving the fits from the 128-channel \rmfit analysis (green) and the 128-channel
MLEfit analysis (gold) of a source position with a $27^\circ$ angle to the detector normal. The deconvolution and reconvolution
assume this same source position.  The purple data point shows the convolution of the fit reported in VC+16
with these same detector responses.  The fit in VC+16, however, is not for this source position but samples the LIGO arc,
with most of the positions with angles closer to $70^\circ$ from the detector normal
of NaI 5.  Using this inconsistent response results
in an overestimation of the predicted counts by a factor of nearly three.  In the right panel we take the 8-channel
\rmfit fit parameters for the position at $27^\circ$ to NaI~5, convolve them with the appropriate response, and show that observed count
rates are matched to the predictions.  The error bars come from sampling the parameter space of the spectral fit to the data
and appear consistent with the counting errors associated with the observations.
This figure also demonstrates how weak GW150914-GBM is in an individual detector, standing out less than a statistical fluctuation
5~s later that disappears when multiple detectors are combined.
 \label{fig:lc}}}
 \end{figure}

Taking a single source position along the LIGO arc, 
JG+16 calculate how many counts above background would be expected in a single detector, NaI 5,
using the central parameter values from the various fits
 and the detector responses for a source at the assumed location. This
should be a simple reversal of the fit, where folding the fit parameters through the detector response
yields the observed count rates above background.  Instead, Figure 5 of JG+16, reproduced in 
the left panel of Figure \ref{fig:lc}, shows that the parameters obtained using \rmfit with 128-channel
data yield an expected count
rate in excess of the observed rate (green data point) whereas the parameters obtained using MLEfit are consistent with
the observed count rates (gold).  This is consistent with their conclusion that in the low-count regime, \rmfit
overestimates the amplitude of a weak source, so that convolving the fit with the detector response results in
expected count rates above those observed. 
The authors also take the spectral fit parameters reported in VC+16
and fold them through the detector responses for the entirely different source position analysed in JG+16.
  The count rates (purple) from this convolution exceed the observed rates, 
being comparable to and even higher than the green point.
The authors conclude that 
the 8-channel fits in VC+16 suffer the same problems as the 128-channel \rmfit analysis in JG+16, and are therefore not reliable.  

%\begin{itemize}
%\item 
$\rightarrow$ The fit parameters in VC+16 were obtained by sampling source positions 
over the entire LIGO arc, with most of the positions contained in a sky region where
the angle of the source to NaI 5 was $\sim 70^\circ$. The convolution of the parameters 
in Figure 5 of JG+16 uses
responses for a source position at $\sim 30^\circ$ to NaI 5. 
The use of this inconsistent and therefore
incorrect detector response results in an overestimation of the expected signal by approximately a factor of three,
consistent with the purple data point in the figure.     
%\end{itemize}

In the right panel of Figure \ref{fig:lc} we show the convolution of the fit parameters obtained using \rmfit with 8-channel
data for the source position assumed by JG+16,
demonstrating that the observed count rates are matched when the fit parameters are convolved
with the appropriate response, as expected. The fit parameters for this source position are not listed in VC+16
and have only a small weight in the overall fit reported in VC+16.
 They contribute only minimally to the uniform sampling over the LIGO arc, with most of the positions on the arc at
far larger angles to NaI~5. 

We do not overplot our predicted counts for this source position 
on the left panel of Figure \ref{fig:lc}
because the convolution produces counts above background.
We were unable
to reproduce the background fit of JG+16 (red line on the left panel of Figure \ref{fig:lc}), which appears to 
be around 1025 counts/sec between 11 and 930~keV. By contrast, the background in VC+16
is around 1010 counts/sec in the energy range 12 to 980~keV.
Background fits can differ according to the time intervals or the model used for the fit.

The counts above background predicted from the fit obtained in the \rmfit analysis of 128-channel data from
JG+16 on the left panel do appear to be much higher than those on the right
 panel from our \rmfit fit for a source at the same position.
 We attribute our ability to match the observations, 
while the \rmfit analysis of JG+16 overestimates the expected counts, 
to our mitigation 
through the use of 8-channel data
of the shortcomings of \rmfit in the low-count regime. 
 
%In Figure 1 of JG+16, the authors suggest
%that even  8-channel background data suffer from low-count statistics in most of the 8 energy channels.
%As stated above,
%the background fit in JG+16
%above which source counts are summed is higher than that of VC+16.  
%In the 8-channel data with background fits from VC+16, we find that the counts in each energy bin 
%exceed xx in all but x channels (I NEED NUMBERS HERE!!!).

\subsection{Use of excluded source location}

We now consider the choice of source position along the LIGO arc used in JG+16 for 
the comparison of the spectral fits obtained using \rmfit and MLEfit. It is the northernmost position
on the southern part of the LIGO arc, listed in 10th position
in Table 2 of VC+16, where the order
is from south to north. The five southernmost points contain
over 50\% of the total probability in the LIGO localization region, and the five
northern points of the southern arc a total of around 20\%, with the position selected in JG+16
containing only 2\% of the LIGO localization probability.  
%(It is placed
%in the second row of Table 1 in JG+16, where the
%rest of the positions on the arc arc listed in the same order as in VC+16.)
This position on the arc is notable in having
the smallest angle to NaI 5  ($27^\circ$) and the most favorable position on the southern arc for GBM in 
general. A fluence calculated for a source at this position will be lower after deconvolution of the
detector responses than for a source at a less favorable geometry to GBM. 

%\begin{itemize}
%\item
$\rightarrow$ It can be seen in Figure \ref{fig:gbm_loc} that the source position 
selected by JG+16
for spectral analysis of GW150914-GBM 
is completely excluded by the GBM localization ($> 3\sigma$ level).
The relative rates measured in the GBM detectors are incompatible with a source at that
location, implying that any spectral fits to the data assuming this source position are not reliable.
%\end{itemize}

Figure \ref{fig:lc} shows that 
even an unreliable fit can be used to recover the observed count rates when the fit parameters
are convolved with the appropriate detector response, both for the MLEfit analysis and the 8-channel \rmfit analysis. 
% This is true both for the MLEfit
% analysis in JG+16 and the \rmfit analysis for the same source position presented here, 
The choice of a source position excluded by the GBM localization could be a problem, however, 
if used to infer other properties of the source.  

In section 4.2 of JG+16, the authors present a Bayesian approach to source detection 
that considers not just the amplitude of the source signal, but also its
spectral distinctness from the background data surrounding the time interval of interest.  
This novel
technique would allow the detection of
 a weaker transient above background if its energy spectrum were sufficiently 
distinct from the spectrum of the background whereas a transient with a similar energy spectrum to the 
background would need to be brighter to be distinguished above the background.

Using the data from two GBM detectors (NaI 5 and BGO 0), JG+16
test whether a source exists above background and find no preference for a source. When they inject
a simulated source of the amplitude and spectral power-law index reported in VC+16 they find a preference
for a source above background. 
The authors attribute this effect to the similarity between the spectrum of GW150914-GBM in the real data and the spectrum of 
the background data around it.  In the simulation, by contrast,
the difference between the source and background data allows the detection of the source.
Their conclusion is that the data in the GW150914-GBM time interval form part of the background.
Conversely, Figure 9 of JG+16 shows that
if a source is injected using responses from a known source position, then as long as the same response
is used in the detection process, the simulated transient will have the correct count
rate ratios for the two detectors and will be recovered as a source above background.  The authors do not show that an injected simulated
transient with the same amplitude but with a spectrum similar to their background spectrum 
is not recovered in their analysis. 

We discuss this Bayesian approach here because it is the only aspect of JG+16 that deals with 
source detection rather than spectral analysis.  The technique could provide 
an alternative approach for signal detection to that detailed in VC+16 with a new dimension that considers not just
the size of a signal and its compatibility with a source based on coherent combination of detector data
but also the distinctness of the putative signal from the surrounding background data.  
The detection of GW150914-GBM presented in VC+16 relies on the combination of
data from all GBM detectors, not just two detectors.
The authors in JG+16 have shown in this introduction to their
 Bayesian approach the ability
to consider data from multiple detectors. A natural next step is
to combine the data from all detectors, which they have done in 
\cite{burgess2017}. In this new paper, they find the transient event
 GW150914-GBM and calculate a fluence similar to that obtained in VC+2016. 
Their assessment laid out in JG+2016
that this is a background fluction is unchanged, based 
on the non-detection of the event by SPI-ACS.  
The crux of VC+2016 is that a transient is found with
a significance, a FAR,
 and a FAP determined empirically without any conclusion as to the
nature of the event itself. The power of the
Bayesian technique presented in JG+16 and 
explained in greater detail in \cite{burgess2017} would be enhanced by
the evaluation of a False Alarm Rate based on running the search on long
 stretches of data. 

\subsection{Detectability of GW150914-GBM by SPI-ACS}

The investigation of the spectral analysis of GW150914-GBM by JG+16 
includes an assessment in section 5 of their paper of the detectability of the GBM transient in 
the anti-coincidence shield of the SPectrometer onboard INTEGRAL (SPI-ACS) \citep{savchenko2016} assuming
spectral fit parameters obtained in the \rmfit and MLEfit 128-channel analysis presented in Table 1.

%\begin{itemize}
%\item
$\rightarrow$ Figure 11 of JG+16 shows that while the detection significances expected in the SPI-ACS
based on the 128-channel \rmfit analysis are very high, the MLEfit analysis suggests fluences compatible with
non-detection by SPI-ACS from all positions on the arc.  The authors do not draw this natural conclusion, 
based on their own spectral analysis.  
%\item

$\rightarrow$ We note that the red crosses in Figure 11 are supposed to 
represent the fit parameters in Table 1 but do not in fact correspond
to those values (Table 1 implies overlapping parameter regions for \rmfit and MLEfit).  
It is not clear from the text which set of
numbers is correct. 
%\item

$\rightarrow$ JG+16 also show in Table 2 the expected fluences 
in the ACS in the 50 -- 4700~keV energy range, using the same list of positions as in Table 1. The 
fluence for their chosen position (line 2 in both tables) is the second highest but
should be among the lowest in the
table given the favorable geometry to GBM. This discrepancy is
suggested also by the low expected detection significance in $\sigma$ in the same table.   
%\end{itemize}

We note that predicting signal strengths in the SPI-ACS
from spectral information obtained from GBM is an active area of collaboration between the two
instrument teams.  JG+16 use a
simple power-law fit to the data from GBM, that is unlikely
to represent the true spectrum of the source, to estimate the expected signal size in SPI-ACS, an instrument with
a different response as a function of energy. Ongoing work suggests the true shape of the spectrum, the asymmetry in
fit parameter uncertainties, and the location of the source all play an important role in establishing consistency
among signals in different instruments.

\section{What is the true spectrum of GW150914-GBM and does this affect its detection significance?}
\begin{figure}
   \centering{
   \hbox{
    \includegraphics[width=3.5in]{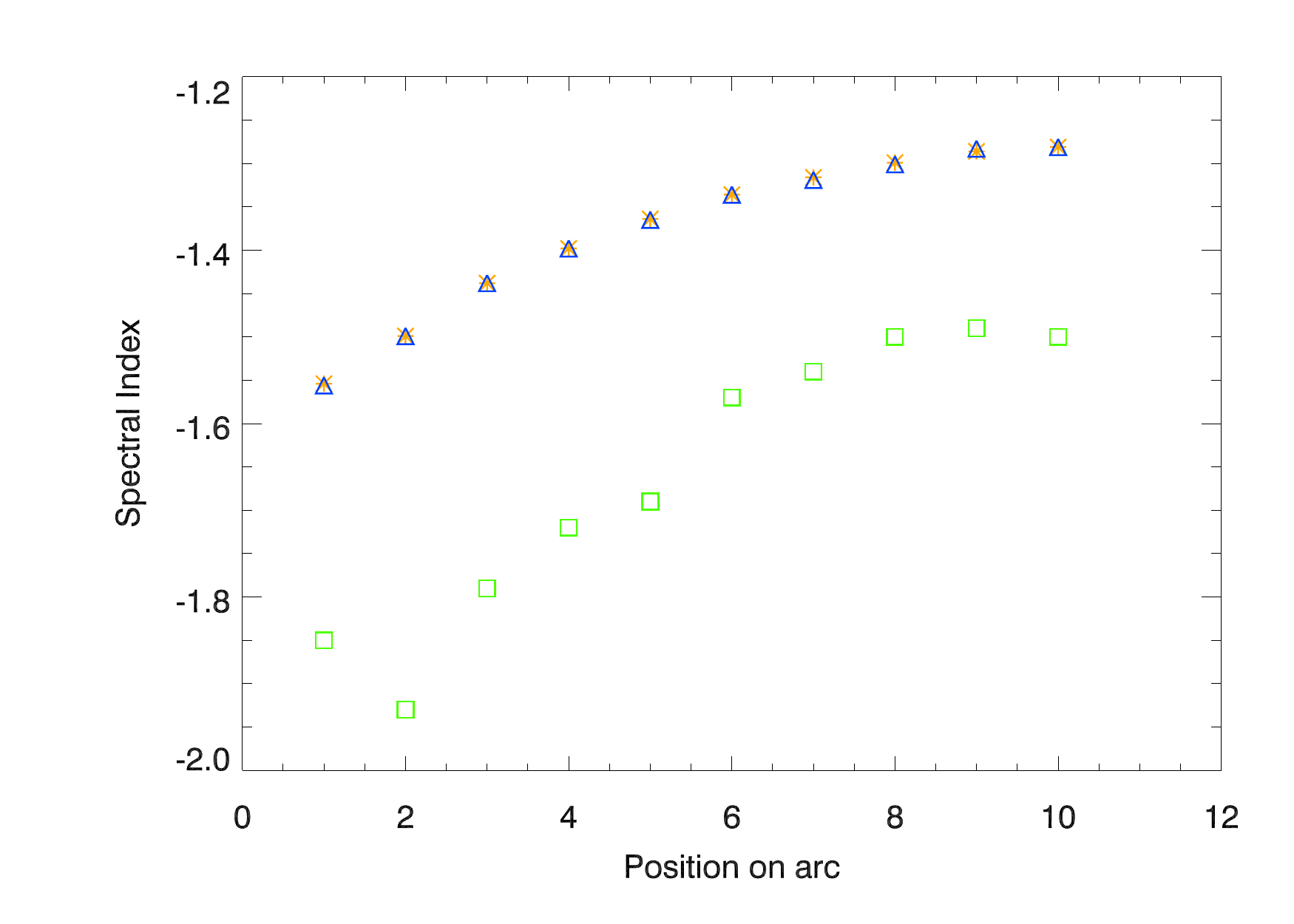}
    \includegraphics[width=3.5in]{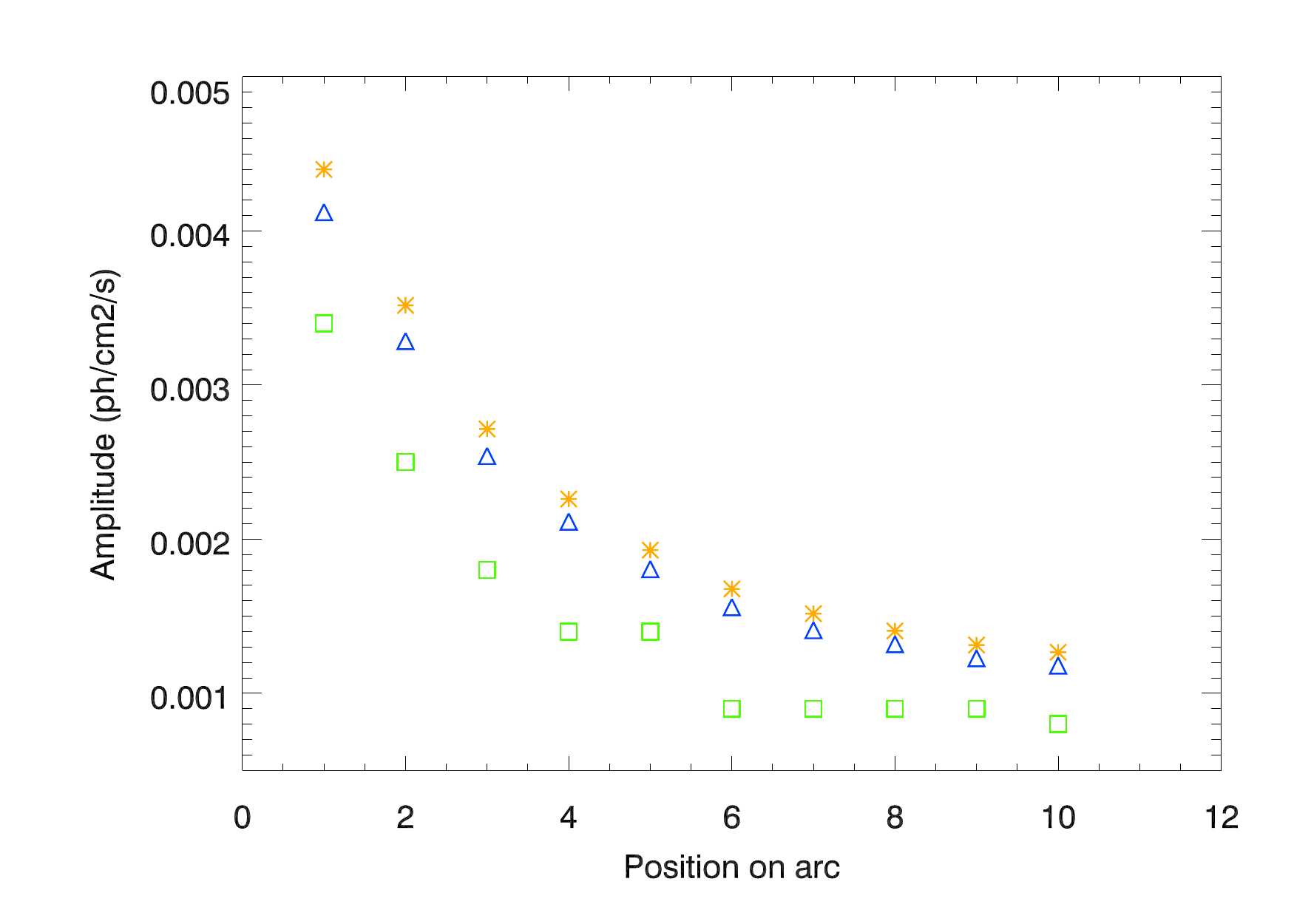}}
    \hbox{
    \includegraphics[width=3.5in]{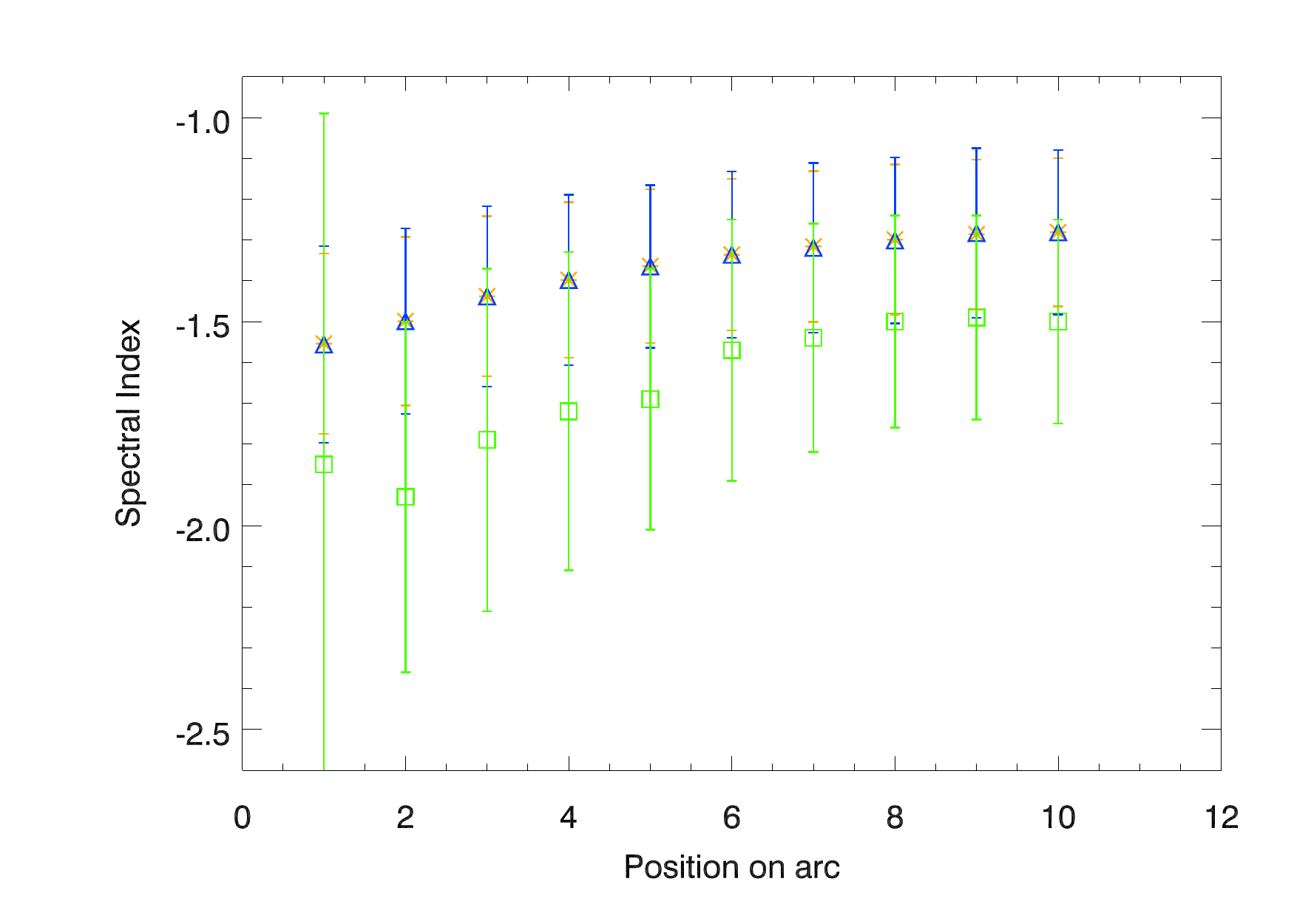}
    \includegraphics[width=3.5in]{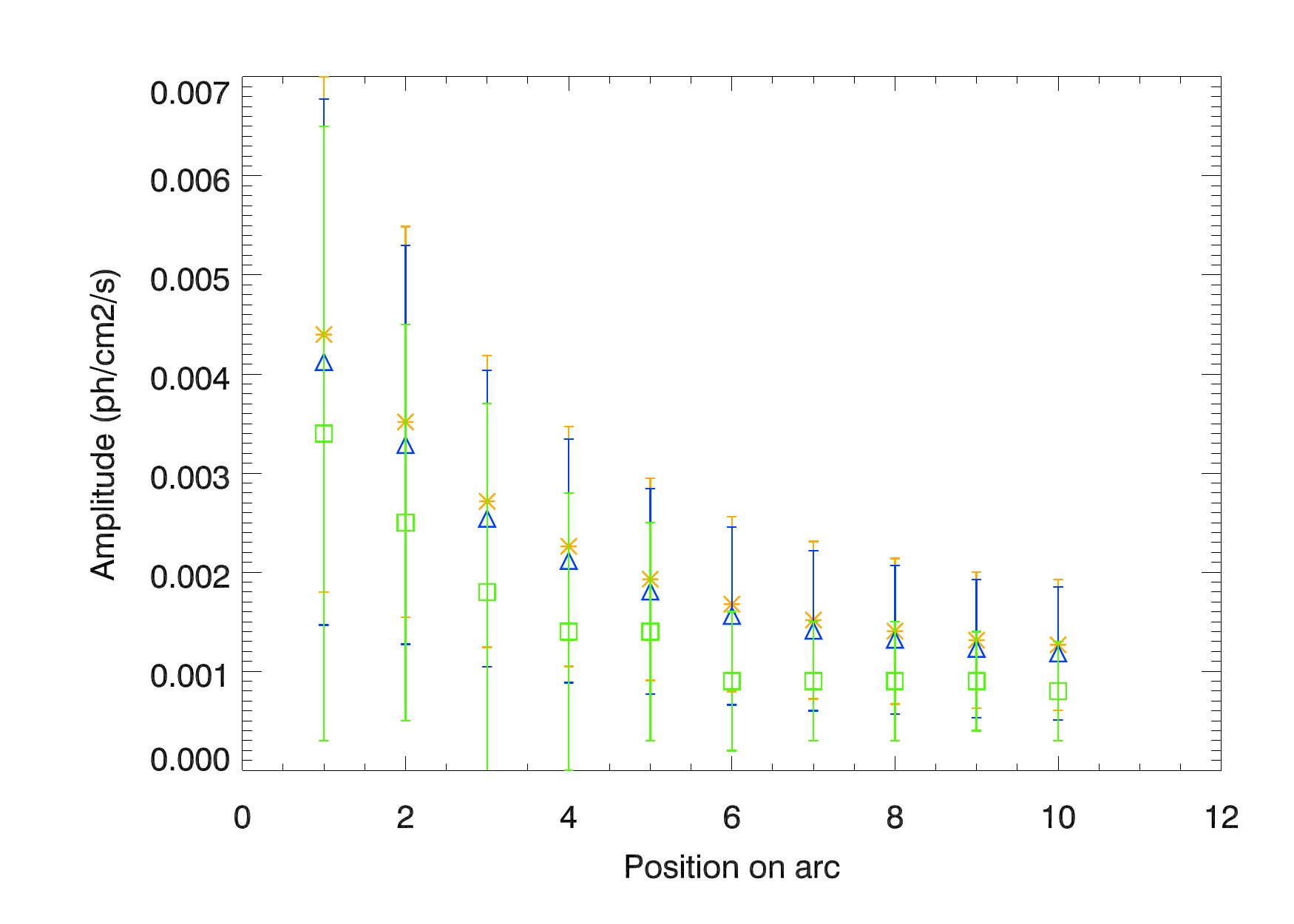}}
    \caption{The left two panels show the variation of the spectral index returned by spectral fits 
to the data from GW150914-GBM as a function of position on the LIGO arc, 
going from south to north on the southern portion of the arc (first 10 points in Table 2 of VC+16).
Parameter uncertainties are excluded from the top panel for clarity but shown in the lower panel. 
The gold stars show the 8-channel
\rmfit fits used in VC+16, the blue triangles an XSPEC analysis of the same data using PGStat as a fitting statistic, and the lime-green squares
are the fits reproduced from Table 1 in JG+16, using 128-channel data and the MLEfit package.  The amplitude of the fit at
100~keV is shown in the right-hand panels.  
In the case of \rmfit and XSPEC, the uncertainties are 68\% confidence level, 1-D errors
returned by the fitting package. For MLEfit they come from a Monte Carlo simulation.  
 \label{fig:spec_comp}}}
 \end{figure}

Spectral analysis of data from weak transients seen in high-background
detectors is limited by our understanding of detector responses, background variations, and modeling of source spectra.  
The tools developed for the analysis of bright transients, where background 
uncertainties do not play a large role, need improvements to deal with low-count statistics.
We have modified \rmfit to deal with background uncertainties in a manner similar to MLEfit and repeated our analysis of
GW150914-GBM with the 8-channel data used in VC+16. 
We do not recover the softer power-law indices, lower amplitudes,
 and large uncertainties reported in JG+16.  We repeated the analysis using XSPEC \citep{arnaud1996}, the
standard spectral fitting tool in the high-energy astronomical 
community\footnote{\url{https://heasarc.gsfc.nasa.gov/xanadu/xspec/manual/XspecManual.html}}.
We use XSPEC version 12.9.1 with the PGStat
fitting statistic, which accounts for non-Poisson background
in Poisson data, overcoming the limitations of \rmfit in its treatment of
background uncertainties in the low-count regime.
For the source position explored in JG+16, we find an amplitude (at 100~keV) 
of $1.18 \pm 0.67 \times 10^{-3}$ ph/cm$^2$/s compared
with $1.27 \pm 0.66 \times 10^{-3}$ using \rmfit and $0.8 \pm 0.5 \times 10^{-3}$ with MLEfit in JG+16.  The indices
returned by \rmfit and XSPEC are very similar ($-1.28 \pm 0.18$ versus $-1.28 \pm 0.20$) with MLEfit producing a much
softer (but still within errors) index of $-1.50 \pm 0.25$.  In Figure \ref{fig:spec_comp} we show the comparison along the
LIGO arc of the three packages for both the amplitude and the index of the fit.  Although XSPEC produces systematically slightly 
lower amplitudes (using the same data and background fits) than \rmfit, the differences are small, much smaller than the differences between
either package and MLEfit (using different background selections).
The indices returned by XSPEC and \rmfit show no systematic deviation and are much harder than those
returned by MLEfit.
Noting that our background
fit means more counts are attributed to the source
 than in JG+16, we suggest that the main difference between the analysis of 128-channel
data with MLEfit and 8-channel data with \rmfit may be systematic -- a consequence of different background fits -- 
instead of statistical.  It is clear that both the
level of the background established by the fit and the statistical uncertainties in the background rates over the source time interval   
play important roles in the analysis and that calculating the true spectrum and fluence of such a weak transient is difficult.

We now consider how the significance of the detection of GW150914-GBM is affected by the possibility that the spectral analysis reported
in VC+16 is flawed and the spectral analysis reported in JG+16 is a better representation of the true spectrum of the
transient. As described above, GW150914-GBM was found in
a coherent search over the whole sky using the data from all 14 GBM detectors in a 60~s time window centered on GW150914. 
At each tested sky position and in each time interval (overlapping bins from 0.256~s to 8.192~s) the full instrument response is convolved
with three template source spectra in turn and the likelihood of a source
being present compared with just background is evaluated.
The search reports the likelihood for the time interval and source template that produces the highest likelihood value.
We note that the background fits for the search are performed automatically and completely independently of the background fits in any
subsequent spectral analysis.
No optimization of spectral fit parameter values occurs during the search procedure
-- just a convolution of the count data with the instrument response using the template spectra -- with no guarantee
that even one
 of the three templates is a good fit to the data.  Including more templates could make the search more sensitive to a particular
transient but at the cost of trials factors when evaluating the FAR.
The three templates are considered a good balance of sensitivity
to a broad range of sources GBM could detect -- 
soft (galactic), normal (typical of long GRBs and some short GRBs), and hard (some short GRBs) -- and avoiding unnecessary trials factors.  
A transient might produce
a low FAR with more than one of the templates 
but the search reports only the result for the template spectrum
yielding the highest likelihood. A different set of template spectra may result in a different preferred spectrum and a different likelihood for
a given candidate but a valid result will be obtained for the FAR
so long as this new set of spectra are used in the search over enough data to evaluate how the likelihood converts to a FAR.   

The bulk of the analysis in JG+16 concerns the spectral fits to the GW150914-GBM data, which play no role in the detection of
the transient or the evaluation of its significance. 

%\begin{itemize}
%\item
$\rightarrow$ The concept of the FAR being an ``optimistic lower limit" to the FAR because the spectral template is not a good representation of 
the true spectrum is
meaningless in the context of an empirically derived FAR, as described above.  The search may not be as sensitive to a source if its spectral
templates do not adequately represent the true source spectrum. This will result in a less sensitive search, not an unreliable FAR.
%\item

$\rightarrow$ The repeated assertion of JG+16 that the spectrum of GW150914-GBM is ``very soft"  
resembling a galactic source is not borne out by
their own fits or by cursory visual inspection of the detector count data.  The soft spectrum in the search turns over above 70~keV to a steep
power-law index of -3.5 but we see a signal in BGO~0 above 200~keV.  
If soft transients found in our search preferred a hard template then we should expect the
search to reveal galactic transients preferring the hard template, which does not happen.  The higher FAR and galactic concentration of 
the candidates preferring the soft spectrum suggests the search efficiently finds these numerous transients with the soft template.
%\item

$\rightarrow$ As stated by JG+16, the discovery lightcurve in VC+16 is indeed model-dependent, and shown only as a visual aid.  
Figure 7 in Appendix C of VC+16 shows a raw count-rate lightcurve with a
signal-to-noise ratio of $6 \sigma$, demonstrating that
the signal-to-noise ratio does not depend strongly on an assumed spectral shape.
%\item

$\rightarrow$ The signal-to-noise ratio of 5.1 in the model-dependent discovery lightcurve
does not relate directly
to the value of $< 3 \sigma$ as suggested by JG+16.  
The value of $2.9 \sigma$ is derived from the calculation of the likelihood that a source is present compared with just background, 
converted into a FAR and a FAP.  It is a post-trials estimate of how likely it is that a transient 
of the size and consistency with a point source indicated by the likelihood and associated FAR
(whether background or astrophysical) occurs by chance
so close in time to a GW event.  
%\end{itemize}

\section{Discussion}

Investigations of the GBM data have been carried out by several different groups since VC+16
 first announced the unexpected potential counterpart of GW150914.
In an analysis by \cite{xiong2016}, the author suggests
that the signal above background is higher in the 10
detectors with a poor viewing geometry to the most likely GW position than in the 4 detectors with a good
viewing geometry. He ignores the detector responses themselves, considering
only the angle of the source to the detector normal.  The second
brightest NaI detector, NaI~9 has the largest angle to the GW position but better sensitivity to the source than
some of the other NaI detectors because of the efficiency of the
detector through its back side.  Similarly, BGO~1 is classed as a detector with a bad viewing geometry but 
both BGO detectors, however,
 have roughly equal exposure to any transient sources with a direction extending underneath the spacecraft.
Thus the separation of detectors into good and bad based on this geometric
factor without considering the response of the detectors, the mass model of the spacecraft including blocked detectors, 
and scattered flux
into detectors from the spacecraft and the Earth is simplistic.  The detection and localization of GW150914-GBM implicitly considers
these geometrical factors when a source is found and localized based on the count rates in all the detectors and the 
full instrumental response.  

Another team \citep{bagoly2016} has analyzed the GBM data
and found potential transient counterparts near the time of GW150914 and also LVT151012, a GW candidate for which our
search uncovers no potential counterpart \citep{racusin2017}. They combine data from multiple detectors to look for evidence of a source but do not
do so coherently with respect to either energy channels or detectors.  Instead they attribute weights to each detector
and energy channel based on the signal to noise in that detector and channel. The weight they attach to each detector and each
energy channel does not require a sensible energy spectrum or a detector combination consistent with a single location on
the sky. We fear that this approach magnifies statistical fluctuations in the case that no source is present in the data.

In the work presented here we concentrate on the analysis of the GBM data for GW150914 by JG+16.    
The bulk of the work presented in JG+16
concerns spectral analysis techniques for background-limited data in the low-count regime,
a welcome discussion of an important topic for the analysis of weak transients we might expect to
find in association with future GW events.
% if indeed GW150914-GBM is related to the GW event GW150914. 
The ability to determine the source
spectrum affects the calculation of the fluence and the detectability of the transient by instruments such as SPI-ACS.  

The comparison of the two statistical approaches in JG+16 that highlights 
the shortcomings of \rmfit in the low-count regime is performed on a data type with 128 energy channels,
while the analysis of VC+16 uses 8-channel 
data. The use
 of 8-channel data mitigates the known limitations of \rmfit when calculating fit parameters and their uncertainties
in the low-count regime.
JG+16 convolve the spectral fit parameters reported in VC+16 with 
detector responses for a source at a much smaller angle to the detector than the source angle from which the fits were
obtained. 
This results in an overestimation of the expected count rates in
that detector by approximately a factor of 3. 
We show in the right panel of Figure 2 that convolving the fit parameters in VC+16 with detector responses for a source from
the position from which the fits were obtained
produces count rates that are consistent with the observations. 

JG+16 obtain spectral fit parameters that differ from those presented in VC+16.  
We implemented a similar treatment of background uncertainties to theirs
in \rmfit but were unable to reproduce the results of JG+16.  
On the other hand, we reproduced the spectral analysis results of VC+16 using the XSPEC fitting package and the PGStat fitting statistic, 
so we speculate that the different fits to the background in the two analyses may be at least partly responsible.
If the background fits of JG+16 are correct, then the lower fluences obtained
by JG+16 are consistent with the non-detection of GW150914-GBM by the SPI-ACS.
Based on the spectral fits to GW150914-GBM reported in VC+16, the non-detection by SPI-ACS is constraining \citep{savchenko2016},
requiring a turnover in the spectrum and/or a lower amplitude.
The spectral analysis of this weak transient is clearly challenging and subject to systematic uncertainties in addition to the 
statistical issues noted in JG+16.
 
JG+16 use the results of their spectral analysis to challenge the significance of GW150914-GBM, which relies on the FAR
associated with its likelihood value.
The FAR is an empirical result emerging from a procedure developed {\it a priori} using months of GBM data and is
independent of and unaffected by subsequent investigations of the data that probe the nature of the
transient.  The detection pipeline for the transient is unrelated to \rmfit or to spectral analysis in general,
and uses an automated background fitting procedure independent of any subsequent spectral analysis.  
There are no free parameters beyond time and duration; 
we marginalize over sky position and the analysis is
simply a convolution with the response of each of the detectors using the three template spectra, and
a calculation of the likelihood a source is present given the observations.
GW150914-GBM was uncovered with a preference
for the hard spectral template. The spectral fits reported in JG+16 
and VC+16 lie somewhere between the hard and normal templates
(not the soft template, as stated in JG+16)  
 although neither the normal nor the hard spectrum is a power-law fit. 
It is very likely that the true spectrum, if astrophysical, 
is not a power-law either, but the transient is too weak for
other spectral shapes to be constrained.
Because the FAR is evaluated using the same search over 
months of data, mis-characterizing the spectrum of any class of transient or background fluctuation using inappropriate templates
will be done equivalently in the background 
and when looking for a GW counterpart.  
The search does not have to be maximally efficient and the spectral
templates do not need to represent accurately the true spectral shape of the source. 

\section{Conclusions}

Nearly a year after the publication of the detection of GW150914-GBM as a potential counterpart to GW150914, we revisit the association 
between the weak GBM transient and the GW.  No further transients were uncovered connected with the other GW and
high-confidence GW candidate detected by LIGO during O1, either by GBM \citep{racusin2017}
or by other instruments taking part in the follow-up campaign.
Alternative investigations of the GBM data for GW150914-GBM have been published.  
In light of these independent analyses, we revisit our original analysis and find no reason to question its validity.
Upper limits to the fluence from the non-detection of GW150914-GBM at higher photon energies were
obtained by SPI-ACS \citep{savchenko2016} and by the 
microcalorimeter on-board the Astrorivelatore Gamma a Immagini Leggero (AGILE), 
\citep{tavani2016}. Ongoing collaborative efforts between the GBM and SPI-ACS teams will
determine whether there exists parameter space in which a detection in one instrument can accommodate a non-detection in the other based on the
spectrum and arrival direction of the transient.  It is clear from the comparison of the spectral analyses in VC+16 and JG+16
that determining the spectrum of such a weak transient in a single instrument is difficult, which
complicates the calculation of expected signals in other instruments.

The build-up to the current observing season of LIGO, O2, saw refinements and
improvements to the targeted search of the GBM data \citep{goldstein2016}
and the full deployment of the untargeted search, with candidates 
online\footnote{\url{https://gcn.gsfc.nasa.gov/fermi_gbm_subthresh_archive.html}}
with candidates promptly reported via the Gamma-ray Coordinates 
Network\footnote{\url{https://gcn.gsfc.nasa.gov/admin/fermi_gbm_subthreshold_announce.txt}}.
Improvements to the targeted search include
the replacement of the hard template with a template more typical of sGRBs in which the power-law index turns over steeply above the peak
energy; an improved background-fitting process; 
the incorporation of information from the LIGO localization in ranking the candidates uncovered by the search; and the production of
lightcurves and sky maps to evaluate quickly any interesting candidates that
can then be distributed to fellow LIGO follow-up observers. 
%When applied to GW150914, the new search recovered the candidate GW150914-GBM
%with a FAR ranging from $3.9 \times 10^{-5}$ to $1.7 \times 10^{-3}$~Hz (there is a new phase shift parameter that results in a $\sim 36$ \%
%difference in the ranking statistic), compared with $1.2 \times 10^{-4}$~Hz using the O1 search technique.

In the absence of confirmation from other instruments or new counterpart candidates to other GW events from merging
black holes in binary systems, the believability of the
association between GW150914-GBM and GW150914 still rests on the FAR and the FAP and the supporting analyses reported
in VC+16 that do not exclude the association.  
The alternative analyses of the GBM data do not confirm or challenge this association. 
Further insight into the possible connection of GW150914-GBM with
 the gravitational-wave event GW150914 will likely have to wait for more observations of similar binary black hole mergers.

\begin{acknowledgments}
The GBM project is supported by NASA.  Support for the German contribution to GBM was provided by the Bundesministerium f\"ur
Bildung und Forschung (BMBF) via the Deutsches
Zentrum f\"ur Luft und Raumfahrt (DLR) under contract number 50 QV 0301.
A.v.K. was supported by the Bundesministeriums f\"ur Wirtschaft und Technologie (BMWi) through DLR grant 50 OG 1101.
JV was supported by STFC grant, ST/K005014/1.
NC acknowledges NSF grant, PHY-1505373.
PS acknowledges NSF grant PHY-1404121.
The authors acknowledge the use of Johannes Buchner's proof-reading tool languagecheck.py.
The authors thank the referee for the difficult job of evaluating the contents of three papers during the reviewing of this
manuscript.
\end{acknowledgments}

\bibliographystyle{apj}
%\bibliography{gbm_ligo}

\begin{thebibliography}{}
\expandafter\ifx\csname natexlab\endcsname\relax\def\natexlab#1{#1}\fi

\bibitem[{{Abbott} {et~al.}(2016{\natexlab{a}}){Abbott}, {Abbott}, {Abbott},
  {Abernathy}, {Acernese}, {Ackley}, {Adams}, {Adams}, {Addesso}, {Adhikari},
  \& et~al.}]{abbott2016b}
{Abbott}, B.~P., {Abbott}, R., {Abbott}, T.~D., {et~al.} 2016{\natexlab{a}},
  Physical Review X, 6, 041015

\bibitem[{{Abbott} {et~al.}(2016{\natexlab{b}}){Abbott}, {Abbott}, {Abbott},
  {Abernathy}, {Acernese}, {Ackley}, {Adams}, {Adams}, {Addesso}, {Adhikari},
  \& et~al.}]{singer2016}
---. 2016{\natexlab{b}}, \apjl, 826, L13

\bibitem[{{Abbott} {et~al.}(2016{\natexlab{c}}){Abbott}, {Abbott}, {Abbott},
  {Abernathy}, {Acernese}, {Ackley}, {Adams}, {Adams}, {Addesso}, {Adhikari},
  \& et~al.}]{abbott2016}
---. 2016{\natexlab{c}}, Physical Review Letters, 116, 061102

\bibitem[{{Abbott} {et~al.}(2016{\natexlab{d}}){Abbott}, {Abbott}, {Abbott},
  {Abernathy}, {Acernese}, {Ackley}, {Adams}, {Adams}, {Addesso}, {Adhikari},
  \& et~al.}]{singer2016b}
---. 2016{\natexlab{d}}, \apjs, 225, 8

\bibitem[{{Arnaud}(1996)}]{arnaud1996}
{Arnaud}, K.~A. 1996, in Astronomical Society of the Pacific Conference Series,
  Vol. 101, Astronomical Data Analysis Software and Systems V, ed. G.~H.
  {Jacoby} \& J.~{Barnes}, 17

\bibitem[{{Bagoly} {et~al.}(2016){Bagoly}, {Sz{\'e}csi}, {Bal{\'a}zs},
  {Csabai}, {Horv{\'a}th}, {Dobos}, {Lichtenberger}, \&
  {T{\'o}th}}]{bagoly2016}
{Bagoly}, Z., {Sz{\'e}csi}, D., {Bal{\'a}zs}, L.~G., {et~al.} 2016, \aap, 593,
  L10

\bibitem[{{Blackburn} {et~al.}(2015){Blackburn}, {Briggs}, {Camp},
  {Christensen}, {Connaughton}, {Jenke}, {Remillard}, \&
  {Veitch}}]{blackburn2015}
{Blackburn}, L., {Briggs}, M.~S., {Camp}, J., {et~al.} 2015, \apjs, 217, 8

\bibitem[{{Blackburn} {et~al.}(2013){Blackburn}, {Briggs}, {Camp},
  {Christensen}, {Connaughton}, {Jenke}, \& {Veitch}}]{blackburn2013}
---. 2013, ArXiv e-prints, arXiv:1303.2174

\bibitem[{{Burgess} {et~al.}(2016){Burgess}, {Yu}, {Greiner}, \&
  {Mortlock}}]{burgess2017}
{Burgess}, J.~M., {Yu}, H.-F., {Greiner}, J., \& {Mortlock}, D.~J. 2016, ArXiv
  e-prints, arXiv:1610.07385

\bibitem[{{Connaughton} {et~al.}(2015){Connaughton}, {Briggs}, {Goldstein},
  {Meegan}, {Paciesas}, {Preece}, {Wilson-Hodge}, {Gibby}, {Greiner}, {Gruber},
  {Jenke}, {Kippen}, {Pelassa}, {Xiong}, {Yu}, {Bhat}, {Burgess}, {Byrne},
  {Fitzpatrick}, {Foley}, {Giles}, {Guiriec}, {van der Horst}, {von Kienlin},
  {McBreen}, {McGlynn}, {Tierney}, \& {Zhang}}]{connaughton2015}
{Connaughton}, V., {Briggs}, M.~S., {Goldstein}, A., {et~al.} 2015, \apjs, 216,
  32

\bibitem[{{Connaughton} {et~al.}(2016){Connaughton}, {Burns}, {Goldstein},
  {Blackburn}, {Briggs}, {Zhang}, {Camp}, {Christensen}, {Hui}, {Jenke},
  {Littenberg}, {McEnery}, {Racusin}, {Shawhan}, {Singer}, {Veitch},
  {Wilson-Hodge}, {Bhat}, {Bissaldi}, {Cleveland}, {Fitzpatrick}, {Giles},
  {Gibby}, {von Kienlin}, {Kippen}, {McBreen}, {Mailyan}, {Meegan}, {Paciesas},
  {Preece}, {Roberts}, {Sparke}, {Stanbro}, {Toelge}, \& {Veres}}]{vc2016}
{Connaughton}, V., {Burns}, E., {Goldstein}, A., {et~al.} 2016, \apjl, 826, L6

\bibitem[{{Goldstein} {et~al.}(2016){Goldstein}, {Burns}, {Hamburg},
  {Connaughton}, {Veres}, {Briggs}, {Hui}, \& {The GBM-LIGO
  Collaboration}}]{goldstein2016}
{Goldstein}, A., {Burns}, E., {Hamburg}, R., {et~al.} 2016, ArXiv e-prints,
  arXiv:1612.02395

\bibitem[{{Greiner} {et~al.}(2016){Greiner}, {Burgess}, {Savchenko}, \&
  {Yu}}]{greiner2016}
{Greiner}, J., {Burgess}, J.~M., {Savchenko}, V., \& {Yu}, H.-F. 2016, \apjl,
  827, L38

\bibitem[{{LIGO Scientific Collaboration} {et~al.}(2015){LIGO Scientific
  Collaboration}, {Aasi}, {Abbott}, {Abbott}, {Abbott}, {Abernathy}, {Ackley},
  {Adams}, {Adams}, {Addesso}, \& et~al.}]{aligo}
{LIGO Scientific Collaboration}, {Aasi}, J., {Abbott}, B.~P., {et~al.} 2015,
  Classical and Quantum Gravity, 32, 074001

\bibitem[{{Meegan} {et~al.}(2009){Meegan}, {Lichti}, {Bhat}, {Bissaldi},
  {Briggs}, {Connaughton}, {Diehl}, {Fishman}, {Greiner}, {Hoover}, {van der
  Horst}, {von Kienlin}, {Kippen}, {Kouveliotou}, {McBreen}, {Paciesas},
  {Preece}, {Steinle}, {Wallace}, {Wilson}, \& {Wilson-Hodge}}]{meegan2009}
{Meegan}, C., {Lichti}, G., {Bhat}, P.~N., {et~al.} 2009, \apj, 702, 791

\bibitem[{{Racusin} {et~al.}(2017){Racusin}, {Burns}, {Goldstein},
  {Connaughton}, {Wilson-Hodge}, {Jenke}, {Blackburn}, {Briggs}, {Broida},
  {Camp}, {Christensen}, {Hui}, {Littenberg}, {Shawhan}, {Singer}, {Veitch},
  {Bhat}, {Cleveland}, {Fitzpatrick}, {Gibby}, {von Kienlin}, {McBreen},
  {Mailyan}, {Meegan}, {Paciesas}, {Preece}, {Roberts}, {Stanbro}, {Veres},
  {Zhang}, {Fermi LAT Collaboration}, {Ackermann}, {Albert}, {Atwood},
  {Axelsson}, {Baldini}, {Ballet}, {Barbiellini}, {Baring}, {Bastieri},
  {Bellazzini}, {Bissaldi}, {Blandford}, {Bloom}, {Bonino}, {Bregeon}, {Bruel},
  {Buson}, {Caliandro}, {Cameron}, \& et~al.}]{racusin2017}
{Racusin}, J.~L., {Burns}, E., {Goldstein}, A., {et~al.} 2017, \apj, 835, 82

\bibitem[{{Savchenko} {et~al.}(2016){Savchenko}, {Ferrigno}, {Mereghetti},
  {Natalucci}, {Bazzano}, {Bozzo}, {Brandt}, {Courvoisier}, {Diehl}, {Hanlon},
  {von Kienlin}, {Kuulkers}, {Laurent}, {Lebrun}, {Roques}, {Ubertini}, \&
  {Weidenspointner}}]{savchenko2016}
{Savchenko}, V., {Ferrigno}, C., {Mereghetti}, S., {et~al.} 2016, \apjl, 820,
  L36

\bibitem[{{Siellez} {et~al.}(2014){Siellez}, {Bo{\"e}r}, \&
  {Gendre}}]{siellez2014}
{Siellez}, K., {Bo{\"e}r}, M., \& {Gendre}, B. 2014, \mnras, 437, 649

\bibitem[{{Tavani} {et~al.}(2016){Tavani}, {Pittori}, {Verrecchia},
  {Bulgarelli}, {Giuliani}, {Donnarumma}, {Argan}, {Trois}, {Lucarelli},
  {Marisaldi}, {Del Monte}, {Evangelista}, {Fioretti}, {Zoli}, {Piano},
  {Munar-Adrover}, {Antonelli}, {Barbiellini}, {Caraveo}, {Cattaneo}, {Costa},
  {Feroci}, {Ferrari}, {Longo}, {Mereghetti}, {Minervini}, {Morselli},
  {Pacciani}, {Pellizzoni}, {Picozza}, {Pilia}, {Rappoldi}, {Sabatini},
  {Vercellone}, {Vittorini}, {Giommi}, {Colafrancesco}, \&
  {Cardillo}}]{tavani2016}
{Tavani}, M., {Pittori}, C., {Verrecchia}, F., {et~al.} 2016, ArXiv e-prints,
  arXiv:1604.00955

\bibitem[{{The LIGO Scientific Collaboration} {et~al.}(2017){The LIGO
  Scientific Collaboration}, {the Virgo Collaboration}, {Abbott}, {Abbott},
  {Abbott}, {Acernese}, {Ackley}, {Adams}, {Adams}, {Addesso}, \&
  et~al.}]{abbott2017}
{The LIGO Scientific Collaboration}, {the Virgo Collaboration}, {Abbott},
  B.~P., {et~al.} 2017, ArXiv e-prints, arXiv:1706.01812

\bibitem[{{Xiong}(2016)}]{xiong2016}
{Xiong}, S. 2016, ArXiv e-prints, arXiv:1605.05447

\end{thebibliography}

\hyphenation{Post-Script Sprin-ger}

\end{document}